\documentclass[twocolumn,A4]{article}
\usepackage[dvips]{graphics}

\begin{document}

\onecolumn

\begin{center}
{\bf{\Large Fractional periodic persistent current in a twisted normal 
metal loop: an exact result}}\\
~\\
Santanu K. Maiti$^{\dag,\ddag,*}$ \\
~\\
{\em $^{\dag}$Theoretical Condensed Matter Physics Division,
Saha Institute of Nuclear Physics \\
1/AF, Bidhannagar, Kolkata-700 064, India \\
$^{\ddag}$Department of Physics, Narasinha Dutt College,
129, Belilious Road, Howrah-711 101, India} \\
~\\
{\bf Abstract}
\end{center}
We explore a novel mechanism for control of periodicity in persistent 
current in a twisted normal metal loop threaded by a magnetic flux $\phi$. 
A simple tight-binding model is used to describe the system. Quite 
interestingly we see that, depending on the number of twist $p$, the 
persistent current exhibits $\phi_0/p$ ($\phi_0=ch/e$, the elementary 
flux-quantum) periodicity. Such fractional periodicity provides a new 
finding in the study of persistent current.
\vskip 1cm
\begin{flushleft}
\end{flushleft}
\vskip 3.5in
\noindent
{\bf ~$^*$Corresponding Author}: Santanu K. Maiti

Electronic mail: santanu.maiti@saha.ac.in
\newpage
\twocolumn

\section{Introduction}

Over the last few decades, the physics at sub-micron length scale provides
enormous evaluation both in terms of our understanding of basic physics
as well as in terms of the development of revolutionary technologies.
In this length scale, the so-called mesoscopic or nanoscopic regime,
several characteristic quantum length scales for the electrons such as
system size and phase coherence length or elastic mean free path and
phase coherence length are comparable. Due to the dominance of the quantum
effects in the mesoscopic/nanoscopic regime, intense research in this 
field has revolved its richness. The most significant issue is probably
the persistent currents in small normal metal rings. In thermodynamic 
equilibrium, a small metallic ring threaded by magnetic flux $\phi$ 
supports a current that does not decay dissipatively even at non-zero 
temperature. It is the well-known phenomenon of persistent current in 
mesoscopic normal metal rings which is a purely quantum mechanical effect 
and gives an obvious demonstration of the Aharonov-Bohm effect.$^1$ 
An electron moving around the ring without entering the region of magnetic 
flux, feels no classical force during its motion. But the quantum state of
this electron is affected by changing the phase of its wave function
due to the presence of the magnetic vector potential $\vec{A}$, related to 
the magnetic field $\vec{B}$ through the relation 
$\vec{B}=\vec{\bigtriangledown} \times \vec{A}$. Accordingly, both the 
thermodynamic and kinetic properties vary with the magnetic flux $\phi$.
The possibility of persistent current was predicted in the very early days 
of quantum mechanics by Hund,$^2$ but their experimental evidences 
came much later only after realization of the mesoscopic systems. 
In $1983$, B\"{uttiker} {\em et al.}$^3$ predicted theoretically that 
persistent current can exist in mesoscopic normal metal rings threaded by 
a magnetic flux $\phi$, even in the presence of impurity. The first 
experimental verification of persistent current in mesoscopic normal metal
ring has been established in the pioneering experiment of Levy {\em et 
al.}$^4$, and later, the existence of the persistent current was further 
confirmed by several experiments.$^{5-8}$ Though the phenomenon of 
persistent current has been addressed quite extensively over the last 
two decades both theoretically$^{9-29}$ as well as experimentally,$^{4-8}$ 
but the controversy between the theory and experiment cannot be 
resolved yet. As illustrative example, the major controversies appear 
in the determinations of (I) the current amplitude, (II) flux-quantum 
periodicities, (III) low-field magnetic susceptibilities, etc. In recent 
works,$^{23-25}$ it has been predicted that the higher order hopping 
integrals, in addition to the nearest-neighbor hopping (NNH) integral, 
have a significant role to enhance the current amplitude (even an order 
of magnitude). The inclusion of the higher order hopping integrals with 
the NNH integral is much more convenient for the description of a system 
rather than the traditional NNH model, since the overlap of the atomic 
orbitals between various neighboring sites are usually non-vanishing. 
In other recent work,$^{26}$ we have focused that the low-field magnetic 
susceptibility can be mentioned exactly only for the one-channel systems 
with fixed number of electrons, while for all other cases it becomes 
random. To grasp the experimental behavior of the persistent current, 
one has to focus attention on the interplay of quantum phase coherence, 
disorder and electron-electron correlation and this is a highly 
challenging problem.

In this present paper, we concentrate ourselves on a strange behavior of
persistent current which is quite different than that of the above 
mentioned issues. This is the appearance of the fractional periodic 
persistent current in a twisted normal metal loop, which is solely 
different from the conventional systems i.e., one-channel rings or 
multi-channel cylinders. Several interesting phenomena of persistent 
currents have already been studied in some particular twisted systems. 
For example, Yakubo {\em et al.}$^{30}$
have described the behavior of persistent currents in a twisted geometry,
so-called the M\"{o}bius strip, and found the distinct flux-periodicity
than that of a ordinary cylindrical sample. In other work, Ferreira
{\em et al.}$^{31}$ have also investigated the topological effect 
on persistent currents in a M\"{o}bius strip and revealed some interesting
new results. Motivated with such kind of systems, in this paper we focus 
our study on the behavior of persistent current in a $p$-fold twisted 
loop geometry (see Fig.~\ref{loop}). Quite interestingly from our study 
we see that, depending on the number of twist $p$, the persistent current 
exhibits $\phi_0/p$ flux-quantum periodicity. This phenomenon is completely
opposite compared to any traditional ring/cylindrical system, where the 
current exhibits only $\phi_0$ flux-quantum periodicity.

We organize this paper specifically as follows. In Section $2$, we describe 
the model and the theoretical description for our calculations. Section $3$ 
focuses the significant results and the discussion. Here we compute the
persistent currents both for the systems with fixed number of electrons and
fixed chemical potential and focus on the basic mechanism for the control of
periodicity of persistent currents in a twisted geometry. Finally, we draw 
our conclusions in Section $4$.

\section{The model and the method}

The schematic representation of a twisted normal metal loop threaded by a 
magnetic flux 
\begin{figure}[ht]
{\centering \resizebox*{5cm}{3.3cm}{\includegraphics{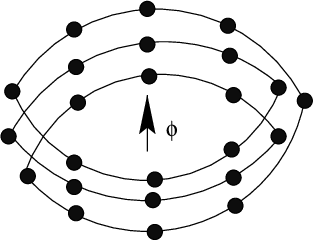}}\par}
\caption{Schematic view of a twisted normal metal loop threaded by a 
magnetic flux $\phi$. The filled circles correspond to the position of 
the atomic sites.}
\label{loop}
\end{figure}
$\phi$ is shown in Fig.~\ref{loop}. The system can be described by a 
single-band tight-binding Hamiltonian and within the non-interacting 
picture, the Hamiltonian looks in this form
\begin{eqnarray}
H = \sum_{i=1}^{pN} \epsilon_i c_i^{\dagger} c_i + v \sum_{<ij>}
\left[e^{i \theta} c_i^{\dagger} c_j+ e^{-i \theta} c_j^{\dagger} 
c_i \right] 
\label{hamil1}
\end{eqnarray}
In this expression, $p$ is the total number of turns and $N$ corresponds to 
the total number of atomic sites in each turn of the loop, $\epsilon_i$'s 
are the site energies, $c_i^{\dagger}$ ($c_i$) is the creation (annihilation) 
operator of an electron at site $i$ of the loop, and $v$ is the hopping
integral between nearest-neighbor sites in each turn. 
Here, $\theta=2\pi \phi/N$ is the phase factor due to the flux $\phi$ 
(measured in units of $\phi_0=ch/e$, the elementary flux quantum). Now in 
order to introduce the impurities in the system, we choose the site 
energies ($\epsilon_i$'s) from an incommensurate potential distribution 
function: $\epsilon_i = W \cos(i\lambda \pi)$, where $W$ is the strength 
of the potential and $\lambda$ is an irrational number, and as a typical 
example we take it as the golden mean $\left(1+\sqrt{5}\right)/2$. Setting 
$\lambda=0$, we get back the pure system with identical site potential $W$. 
The idea of considering such an incommensurate potential rather than the 
usual random distribution is that, for such a correlated disorder we do 
not require any configuration averaging and therefore the numerical 
calculations can be done in the low cost of time. 

At absolute zero temperature, the persistent currents in the loop threaded
by a flux $\phi$ is determined by$^{10}$
\begin{equation}
I(\phi) = - \frac{\partial{E(\phi})}{\partial{\phi}}
\label{curr}
\end{equation}
where, $E(\phi)$ is the ground state energy. We evaluate this energy 
exactly to understand unambiguously the anomalous behavior of persistent 
current, and this is achieved by exact diagonalization of the tight-binding 
Hamiltonian Eq.~(\ref{hamil1}). Throughout the calculations, we take
$v=-1$ and use the units where $c=1$, $e=1$ and $h=1$.

\section{Results and discussion}

\begin{figure}[ht]
{\centering \resizebox*{8cm}{10cm}{\includegraphics{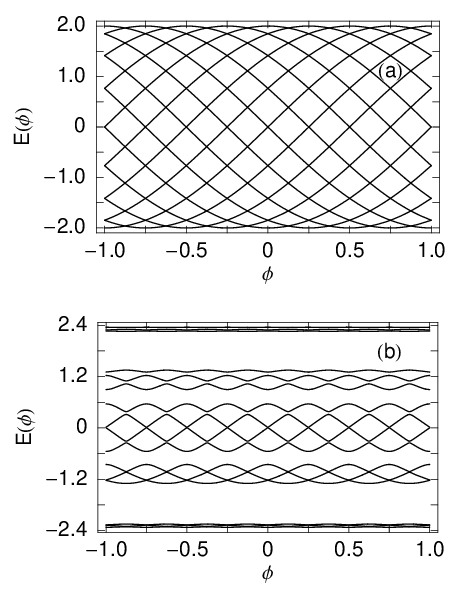}}\par}
\caption{$E(\phi)$-$\phi$ curves for (a) ordered ($W=0$) and (b) disordered 
($W=1$) $4$-fold ($p=4$) twisted loops with total number of atomic sites 
in each turn $N=4$.}
\label{energy}
\end{figure}

In this section, we investigate the behavior of persistent currents both
for the ordered and disordered twisted loops and illustrate how the 
periodicity of the current can be controlled by tuning the total number
of twist in such a particular geometry. To have a deeper insight to the 
problem, let us first start our discussion with the energy-flux 
characteristics of a twisted loop. In an ordered system, we put
$\epsilon_i=0$ for all $i$ in the above Hamiltonian given by 
Eq.~(\ref{hamil1}), and the energy of the $n$-th single-particle state 
can be calculated analytically which is written as
\begin{equation}
E_n(\phi)=2 v \cos\left[\frac{2\pi}{pN}(n+p\phi)\right]
\end{equation}
where $n$ is an integer and restricted in the range: $-pN/2 \leq n < pN/2$. 
For the disordered case, since the energy $E_n$ cannot be done analytically, 
we evaluate it by exact numerical diagonalization of the tight-binding
Hamiltonian (Eq.~(\ref{hamil1})). As illustrative example, in 
Fig.~\ref{energy} we present the energy levels of $4$-fold ($p=4$)
twisted loops, where we fix the total number of atomic sites $N=4$ in 
each turn. Figures~\ref{energy}(a) and (b) correspond to the ordered 
and dirty systems respectively. In the absence of any impurity i.e., 
for perfect loop, the energy levels intersect with each other at 
$\phi=0$ or at the integer multiple of $\phi_0/p$. This indicates that 
the energy levels have degeneracy at these respective values of the 
magnetic flux $\phi$. On the other hand, all these degeneracies move 
out as long as the impurities are given in the system, and gaps open 
at the points of intersection. Accordingly, the energy levels vary 
continuously with respect to the flux $\phi$ (see the curves in 
Fig.~\ref{energy}(b)). Both these intersecting behavior and the 
continuous variation of the energy levels with $\phi$ are quite similar 
in nature as observed in traditional one-channel rings or multi-channel 
cylinders.$^{26}$ But the significant feature that is observed from 
the energy spectra is that, the energy levels show $\phi_0/4$ flux-quantum
periodicity, instead of $\phi_0$. This $\phi_0$ periodicity is observed 
only for the untwisted loops. One can also get the energy levels with 
other fractional periodicity by tuning the total number of twist $p$ in such
a twisted loop. This phenomenon is really very interesting, and since the 
energy levels are $\phi_0/p$ periodic in $\phi$, the persistent current 
will also exhibit the $\phi_0/p$ oscillating behavior, which we will describe 
now. In the forthcoming sub-sections, we present some analytical as well as
numerical calculations and study the behavior of persistent currents $I$ 
in twisted geometries as a function of the flux $\phi$, system size $L$, 
total number of electrons $N_e$, chemical potential $\mu$, total number of
twist $p$ and the strength of disorder $W$.

\subsection{System with fixed number of electrons $N_e$}

\begin{figure}[ht]
{\centering \resizebox*{8cm}{10cm}{\includegraphics{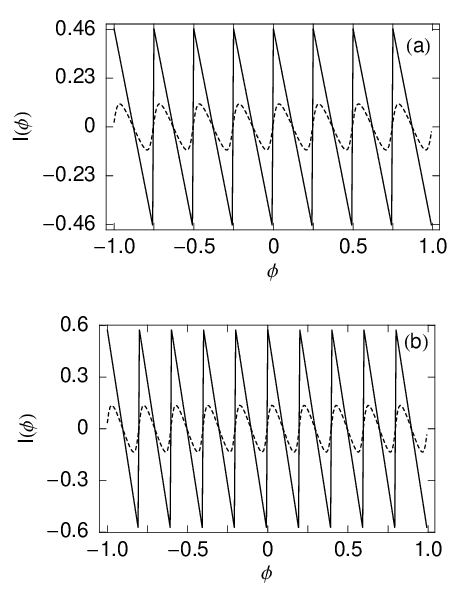}}\par}
\caption{$I(\phi)$-$\phi$ curves for (a) $4$-fold ($p=4$) and (b) $5$-fold
($p=5$) twisted loops with fixed number of electrons $N_e=40$. The total 
number of atomic sites for both the two cases is $100$, where we set
$N=25$ in (a) and $N=20$ in (b). The solid and dotted curves correspond
to the perfect ($W=0$) and disordered ($W=1$) systems respectively.}
\label{current1}
\end{figure}

In the absence of any impurity, the current carried by the $n$-th 
energy eigenstate is
\begin{equation}
I_n(\phi)=\left(\frac{4\pi v}{N}\right)\sin\left[\frac{2\pi}{pN}
(n+p\phi)\right]
\end{equation}
At absolute zero temperature, we can write the total persistent current in
the following form
\begin{equation}
I(\phi)=\sum_n I_n(\phi)
\end{equation}
For the systems with constant $N_e$, the total current will be obtained
by taking the sum of the individual contributions from the lowest $N_e$
energy levels. The variation of the persistent currents with flux $\phi$
for some typical twisted loops is represented in Fig.~\ref{current1}.
Figures~\ref{current1}(a) and (b) correspond to the results 
for the $4$-fold and $5$-fold twisted loops respectively, where the total 
number of atomic sites in both the two cases is taken as $100$. The currents 
are determined for the fixed number of electrons $N_e=40$, where the solid 
and dotted curves represent the currents for the perfect ($W=0$) and 
disordered ($W=1$) systems respectively. From the results it is observed 
that, in the ordered loops the current exhibits the saw-tooth like nature, 
while it shows the continuous variation with much reduced amplitude in 
the disordered cases.
These sharp transitions in the persistent currents for the ordered systems
at the different values of $\phi$ appear due to the degeneracy of the 
energy eigenstates at these respective points. On the other hand, for the 
disordered systems since there is no degeneracy in the energy levels, 
the current varies continuously with the flux $\phi$. Now for these 
disordered systems, we get much reduced current amplitude compared to the 
perfect cases and this is due to the localization effect of the energy 
eigenstates.$^{32}$ Traditional wisdom is that, the larger the 
disorder stronger the localization$^{32}$ and accordingly, we will 
get lesser and lesser current amplitude with the increase of the 
disorder strength $W$. These phenomena are well established in the 
literature. But the most remarkable feature that is observed from 
these current-flux characteristics is that, the periodicity of the 
current changes as we change the total number of twist $p$. Our results 
show that for a fixed system size (here the total number of atomic 
sites is $100$ for both the two cases), the periodicity of the current 
changes from $\phi_0/4$ (see Fig.~\ref{current1}(a)) to $\phi_0/5$
(see Fig.~\ref{current1}(b)) as we change the number of twist from $p=4$ to
$p=5$. This phenomenon can be explained as follows. For a $p$-fold twisted
loop, once an electron starts its motion along the loop from a point, it 
comes back to its original position after traversing the $p$-th turn and
accordingly, it encloses $p\phi$ flux. This originates the $\phi_0/p$ 
oscillations in the persistent currents, instead of $\phi_0$ oscillations
as observed in conventional untwisted loops.$^{26}$ Thus by tuning the 
total number of twist, we can control the periodicity of persistent current. 
This really provides an interesting finding in this particular study.

\subsection{System with fixed chemical potential $\mu$}

\begin{figure}[ht]
{\centering \resizebox*{8cm}{10cm}{\includegraphics{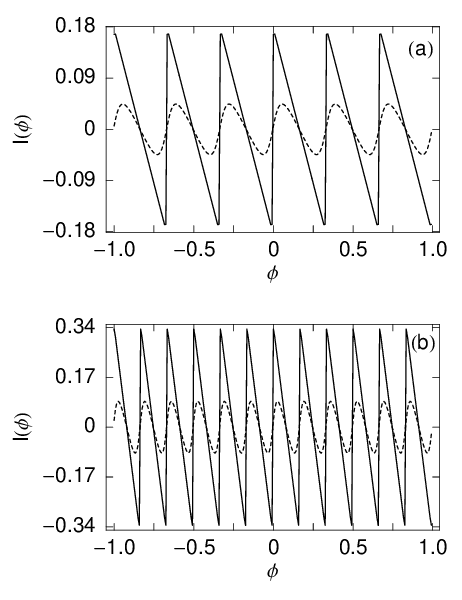}}\par}
\caption{$I(\phi)$-$\phi$ curves for (a) $3$-fold ($p=3$) and (b) $6$-fold
($p=6$) twisted loops with fixed chemical potential $\mu=-1.0$. The total 
number of atomic sites for both the two cases is $180$, where we set
$N=60$ in (a) and $N=30$ in (b). The solid and dotted curves correspond
to the perfect ($W=0$) and disordered ($W=1$) systems respectively.}
\label{current2}
\end{figure}

Now we discuss the characteristic features of the persistent currents for
some twisted loops those are described with fixed chemical potential $\mu$, 
instead of the fixed number of electrons $N_e$. For certain choices of $\mu$, 
the system will have a fixed number of electrons, while for all other cases 
the number of electrons will vary as a function of $\phi$. At absolute zero 
temperature, the total current for such systems (systems with fixed $\mu$)
will be obtained by adding the individual contributions from the energy 
levels with energies less than or equal to $\mu$. As illustrative
example, in Fig.~\ref{current2} we plot the current-flux characteristics
for some $3$-fold and $6$-fold twisted loops respectively, where the total
number of atomic sites in both these two cases is taken as $180$. All the
currents are computed for the fixed chemical potential $\mu=-1.0$, where
the solid and dotted curves represent the same meaning as in 
Fig.~\ref{current1}. Similar to the previous systems, here we also get the
saw-tooth like nature in the persistent currents for the perfect loops, 
while they vary continuously with much reduced amplitude in the presence 
of impurity. Our results clearly show that for a fixed system size (here 
the total number of atomic sites is $180$ for both the two cases), the 
current changes its periodicity from $\phi_0/3$ (see Fig.~\ref{current2}(a)) 
to $\phi_0/6$ (see Fig.~\ref{current2}(b)) as we tune the number of twist 
from $p=3$ to $p=6$. The explanation of this behavior is exactly similar
to that as mentioned earlier. Thus we can emphasize that, the periodicity 
of the persistent currents can be controlled by tuning the total number of
twist in these geometries, whether they are described with fixed $N_e$ or
$\mu$.

\section{Concluding remarks}

In conclusion, we have established a novel feature for control of the 
periodicity of persistent current in a twisted normal metal loop threaded by 
a magnetic flux $\phi$. We have used a simple tight-binding model to describe
the system and provided some analytical as well as numerical calculations
to characterize the current-flux characteristics. From our results 
it has been observed that, the persistent current exhibits $\phi_0/p$
periodicity in a $p$-fold twisted loop, and this periodicity of the current 
can be controlled nicely by tuning the total number of twist in the loop.
Such a peculiar behavior is completely opposite to that of the conventional 
one-channel rings or multi-channel cylinders, where the persistent current
always exhibits $\phi_0$ flux-quantum periodicity. Our theoretical 
results in this article might be helpful to illuminate some of the unusual 
phenomena of persistent currents which have been observed in the twisted
geometries.

Throughout our investigation, we have neglected the effect of 
electron-electron (e-e) correlation since the inclusion of the e-e 
correlation will not provide any new significant result in our present 
study.

\end{document}